
\tolerance 1500
\topskip .75truein

\font\twelverm =cmr12

\def\tt{t^{ons}}
\def\zz{\langle g\rangle}
\def\qs{\lower 5pt\hbox to 23pt{\rightarrowfill}\atop
       {s\rarrow\infty}}
\def\qq{\lower 5pt\hbox to 23pt{\rightarrowfill}\atop
      {q\rarrow\infty}}
\def\q|s|{\lower 5pt\hbox to 23pt{\rightarrowfill}\atop
        {|s|\rarrow\infty}}
\def\ton{\lower 5pt\hbox to 23pt{\rightarrowfill}\atop
       {t\rarrow \tt}}
\def\gon{\lower 5pt\hbox to 23pt{\rightarrowfill}\atop
       {g\rarrow \zz}}

\def\l{\left (}
\def\r{\right )}
\def\o{\over}
\twelverm
\newdimen\offdimen
\def\offset#1#2{\offdimen #1
   \noindent \hangindent \offdimen
   \hbox to \offdimen{#2\hfil}\ignorespaces}
\parskip 0pt

\def\({\lbrack}
\def\){\rbrack}

\def\Oscr{{\cal O}}
\def\Fscr{{\cal F}}

\baselineskip 24pt

\def\aq{\lower 5pt\hbox to 50pt{\rightarrowfill}\atop
       {\hfill a\rarrow\scriptscriptstyle\infty\hfill}}
\def\rin{\lower 5pt\hbox to 50pt{{\rightarrowfill}
     \atop{\hfill r\rarrow\scriptscriptstyle\infty\hfill}}}

\rightline{INFN-ISS 93/2}
\rightline{WIS-93/50/Jun-PH}
\bigskip
\par
\par
\def\tindent#1{\indent\llap{#1}\ignorespaces}
\def\refn{\par\hang\tindent}
\parskip 0pt
\centerline{\bf Exact response of the non-relativistic}
\centerline{\bf harmonic oscillator$^{\dagger}$}
\bigskip
\centerline {E. Pace$^{1}$, G. Salm\`e$^2$}
\centerline {$^1$ Dipartimento di Fisica, Universit\`a di
Roma "Tor Vergata", INFN, Sezione Tor Vergata,}
\centerline {Via della Ricerca Scientifica, I-00133, Roma, Italia}
\centerline {$^2$INFN, Sezione Sanit\`a, Viale Regina Elena 299,
I-00161 Roma, Italy} \centerline {A.S. Rinat}
\centerline {Weizmann Institute of Science, Rehovot 76100, Israel}
\bigskip
\par
\bigskip
\par
\vskip 1truein\noindent
\baselineskip 21pt
\par
{\bf Abstract:}

Using  Green$'$s  function  and  operator techniques  we  give  a  closed
expression  for the  response  of a  non-relativistic system  interacting
through confining,  harmonic forces.   The expression for  the incoherent
part  permits  rapid  evaluation  of coefficients in  a
$1/q$ expansion.  A comparison is made with standard approximation methods.

\vskip.8truein\baselineskip 12pt

\vfil

$^{\dagger}$ This note, together with his best wishes are dedicated by
ASR to his colleague and friend Klaus Dietrich at the occasion of his
60th birthday.
\eject\
\baselineskip 24pt

{\bf 1. Introduction}
\par

Consider a system with mass $m_A$
composed of $A$ constituents with equal
mass  $m$. We  are  interested  in its  linear  response  or  structure
function,  when the  system is  probed  at momentum  and energy  transfer
$(q,\omega)$
$$S(q,\omega)=A^{-1}\sum_n|\Fscr_{0n}(q)|^2
\delta(\omega-q^2/2m_A-E_{n0}),\eqno(1.1)$$
where $\Fscr_{0n}(q)=\langle 0|\rho_q^{\dagger}|n \rangle$ is
the inelastic form factor between the ground state $|0\rangle$
and  excited states $|n\rangle$, and
$\rho_q=\sum_j e_j e^{i\vec{q} \cdot \vec{r}'_j}$
the intrinsic charge density.
$E_{n0}$ are  intrinsic excitation energies
without  target recoil, and $\vec{ r'}_j = \vec{r}_j - \vec{R}$
intrinsic
coordinates with $\vec{R}$ the centre of mass coordinate.

A formal summation over $n$ yields an alternative expression
$$\eqalign{S(q,\omega)&=(\pi A)^{-1}\sum_{i,j=1..A}
{\rm{Im}}\left( \langle 0|e^{-i\vec q \cdot \vec r'_i}G(\omega+E_0-q^2/2m_A
-i\eta) e^{i\vec q \cdot \vec r'_j}|0\rangle\right) \cr
G(z)&=(z-K-V)^{-1}}\eqno(1.2)$$
\noindent
with $G$, the exact Green$'$s function of the system
in terms of the total kinetic and potential energy.

Over the years,  general
methods have been devised to  approximately calculate the
above response.   Recently there has been interest in the  high-$q$
response  for  systems  with  singular  attractive  interactions  of  the
confining  type$^{1,2}$.   For  those systems,  perturbative  methods  in
the confining interaction $V$  fail and non-perturbative methods are required.
We first  mention the  $1/q$ expansion of  the (reduced)  response as
discussed by Gersch, Rodriguez and Smith (GRS)$^3$
$$\phi(q,y)=(q/m)S(q,\omega)=\sum_k(m/q)^kF_k(y),\eqno(1.3)$$
where the energy loss $\omega$
has been replaced by a suitable alternative variable
$$y=-q/2+m\omega/q\eqno(1.4)$$
Alternatively one may perform an actual summation of the series (1.1)
$^2$.  A  pre-requisite for such a treatment is  the knowledge  of
all the states of  the complete spectrum,  which is purely discrete  for
confining interactions.

None of the above methods leads to a closed expression for $S$
or $\phi$.
However, Eq. (1.2) shows that such an expression would result if the
exact Green$'$s function there were known. This is only the case for
a few selected systems. An enumeration of cases, which can be solved
by means of path integral
methods have recently been given $^4$. At this point we recall that
the above method has  been previously suggested  and
applied to the response of systems, as complicated as liquid $^4$He
probed at high $q$ for fixed $y\,\,^5$. Evaluations in practice
will always require approximations.

In what follows, we will  limit ourselves to a system of
particles interacting through harmonic confining forces, which can be
exactly solved and for which approximations can be tested.
Since for this system the
Hamiltonian separates in Jacobi variables,
it suffices to treat only a $'$di-quark$'$  and actually for the
same reason, one in one dimension only. In the following, the mass $m_A$ will
be
replaced by $2m$.

In Section 2  we will
discuss two methods which lead to a closed
expression for the reduced response for the above system. In Section
3 we will give its $1/q$ expansion and compare results with the outcome
of the GRS theory as well as with previously suggested approximations.
\par
{\bf 2. The response for a harmonically confined $'$di-quark$'$.}
\par
{\bf 2a. The Green$'$s function method.}
\par
Consider the Hamiltonian of the relative motion of
a harmonically bound
$'$di-quark$'$
$$H=\hat{p}^2/m+\beta^4\hat{x}^2/m,\eqno(2.1)$$
\noindent
with
$\beta$, the inverse harmonic oscillator
length. For instance the path-integral method leads to the following
expression for the Greens' function for the relative motion
 $^6$.
$$G(x,x',t)=\sqrt{\beta^2\over{2\pi i{\rm sin}\alpha}}{\rm exp}
\left\{(i\beta^2/{2\rm sin}\alpha)\left [(x^2+x'^2)
{\rm cos}\alpha-2xx'\right ] \right\},\eqno(2.2)$$
\noindent
where $\alpha= 2\beta^2s/q\,\,;~s=tq/m$. The ground state wave function is
$\langle x|0\rangle=\(\beta^2/\pi\)^{1/4}e^{-\beta^2x^2/2}$
and the part of the charge density fluctuation dependent
on the relative coordinate
$\rho_q(x)=e_1e^{iqx/2}+e_2e^{-iqx/2}$.
Replacing the relative coordinates $x,x'$ by
$$x=Z+z/2,~~~~~~~ x'=Z-z/2,\eqno(2.3)$$
one finds
$$\rho_q^{\dagger}(x)\rho_q(x')~=~ e_1^2e^{-iqz/2}+e_2^2e^{iqz/2}+
e_1e_2\left( e^{iqZ}+e^{-iqZ}\right)\eqno(2.4)$$
The first two terms are (for equal charges identical)
incoherent contributions due to a transfered $q$, which is absorbed and
emitted by the same particle. For the
remaining coherent contributions those particles are different. The same
transformation (2.3) (which has nothing to do with a CM transformation,
and is  typical for the  harmonic oscillator case) permits  an easy
evaluation of the  $Z,z$ integrals.
Taking the  Fourier  transform of  (2.2)
for the argument $\omega +  \beta^2/m -q^2/4m$, as required
in Eq. (1.2)  and using (1.4) to eliminate $\omega$ in favour of $y$,
one shows $$\eqalignno{\phi&=\phi^{incoh}+\phi^{coh}& \cr \cr
\phi(q,y)^{incoh}&={(e_1^2+e_2^2)\over2}
\int_{-\infty}^{\infty}{ds\over{ 2\pi}}
{\rm exp}\(iys\){\rm exp}\left\(-{\beta^2s^2\over 4}
\left({{\rm sin}\alpha/2\over{\alpha/2}}\right)^2\right\)
{\rm exp}\left\(-iqs/4
\left({{\rm sin}\alpha\over{\alpha}}-1\right)\right\)& (2.5a) \cr \cr
\phi(q,y)^{coh}&=e_1e_2\int_{-\infty}^{\infty}{ds\over{2\pi}}
{\rm exp}\(iys\){\rm exp}\left\({-\beta^2s^2\over 4}
\left({{\rm cos}\alpha/2\over{\alpha/2}}\right)^2\right\)
{\rm exp}\left\(iqs/4\left({{\rm sin}\alpha\over{\alpha}}
+1\right)\right\) & (2.5b)\cr}$$

Except for the harmonic oscillator, we  do not know of  a system for
which the response has been given in a simple closed form like (2.5).  As
is the case  with the GRS theory,  the incoherent part (2.5a)  is seen to
permit a  series expansion  in $\alpha\propto q^{-1}$.  However,
the  GRS expression  $^3$ comprises  three integrals  and in  addition an
ordering operator which can only be realized in the expansion (1.3).

\par
{\bf 2b. Derivation using operator techniques}
\par

In what follows we will sketch a different approach to obtain the response
$S(q,\omega)$, Eq. (1.1), for the case of a $'$di-quark$'$. We do so, by
directly evaluating the time-depending operators after
using the Fourier transform of
the $\delta$ function. Eq. (1.1)  then becomes

$$S(q,\omega)\;=\;{ 1 \over A }\int {dt\over  2 \pi}\;  \langle
0|\rho_q^{\dagger} (\hat{x})\; {\rm exp}({-i Ht})\;\rho_q(\hat{x}) \;{\rm exp}
\left[{i (\omega  - {q^2\over{2m_A}}+ H )t}\right]|0\rangle  \eqno(2.6)$$

Consider first the time-translated
space coordinate $\hat{x}(t)$ i.e.

 $$\hat{x}(t) = {e}^{-iHt}~\hat{x}~
{e}^{iHt}=\sum\nolimits\limits_{i=1}^{\infty
} {t^{n} \over n!}{d{^n}\hat{x}(t) \over dt{^n}}\; \Biggl |_{t=0} \eqno(2.7)$$
where

$${ d{}^{n}\hat{x}(t) \over dt{^n}}\; \Biggl
|_{t=0}=(-i)^n{\left[\right.}\underbrace{{H,\left[H\right.,\left[H\right.}
,.......{\left[H,\right.}}_{n\; times\;H}
{\hat{x}\left.\left.\left.\left.\right]\right]\right]\right]}\eqno(2.8)$$

Using elementary commutation rules and the hamiltonian, Eq. (2.1),
for  the relative motion, one obtains

$$ \hat{x}(t)=\hat{x}\;{\rm cos}(2{{\beta }^{2}{t\over m}})-{\hat{p}\over
{\beta }^{2}} {\rm sin}(2{{\beta }^{2}{t\over m}}) \eqno(2.9)$$

We now apply the Glauber
formula   $e^{A+B} = e^A e^B e^{-{1\over  {2}} [A,B]}$, which
holds if $[A,B]$ is a c-number$^7$, to the operator
${\rm exp}(iq \hat{x}(t)/2)$

$${\rm exp}(iq{\hat{x}(t)\over 2})={\rm exp}\left\{ i{q\over2}
\left[ \hat{x}\;{\rm cos}(2{\beta}^{2}{t\over m})-{\hat{p}\over {\beta
}^{2}}\;{\rm sin}(2{\beta }^{2} {t\over m}) \right] \right \}$$
$$ = {\rm exp}\left\({i q {\hat{x}\over 2}
{\rm cos}(2{{\beta }^{2} {t\over m}})}\right\)\;\; {\rm exp}\left\(
{-iq {\hat{p}\over {2 \beta^{2}}}
{\rm sin}(2{{\beta }^{2} {t\over m}})}\right\)\;\;
{\rm exp}\left[{-i {q^2\over{8\beta^2}}{\rm cos}(2{{\beta }^{2}
{t\over m}}) {\rm sin}(2 {\beta ^2 {t\over m}} ) }\right]
\eqno(2.10)$$

Inserting in Eq.(2.6) the unity operator $
\int dk/(2\pi) |k\rangle \langle k| \;=$ I, and replacing $ m_A$
by $2m$, one obtains
$$S(\omega,q)\;=\;{ 1 \over A }\int {dt\over  2 \pi}\;
{\rm exp}\left[i(\omega  -
{q^2\over{4m}})t\right] \int {dk\over{2 \pi}}  \langle 0|\rho_q^{\dagger}
(\hat{x}) \;{\rm exp}({-i Ht})\;\rho_q(\hat{x}) \;{\rm exp}(i H
t)\;|k\rangle\;\langle k |0\rangle =$$
$$=  \int {dt\over{2 \pi}} {\rm exp}\left[i(\omega -
{q^2\over{4m}})t \right] {\rm exp}\left[-i {q^2\over{16 \beta^2}} {\rm
sin}(4\beta ^{2} {t \over m})\right]\;
\int {dk\over{2 \pi}} \left[{{(e_1^2 + e_2^2)\over2}\; \langle0|k +
{q\over 2}
[ cos(2{\beta ^{2} {t\over m}}) - 1]\;\rangle \; + }\right.$$
$$\left.{ + \;e_1 e_2\; \langle0|k + {q\over 2}
[ {\rm cos}(2{{\beta }^{2} {t\over m}}) + 1]\;\rangle}
\right]\;  {\rm exp}\left[-i {qk\over{2 \beta^2}}
{\rm sin}(2{\beta^2 {t\over m}})\right]\;
\langle k|0\rangle \eqno(2.11)$$

After Fourier transforming the ground state wave function one can easily
perform the integration over $k$. Replacing
the variables $t$ and $\omega$ by
$s$ and $y$, respectively, one obtains Eqs. (2.5a) and (2.5b).

The approach of this section can be extended to any hamiltonian for which
explicit  expressions for  (2.7)  and the  initial  state are  available.
Examples are  the harmonic  oscillator in any  excited initial  state and
many body-systems interacting through harmonic oscillator
forces.

 \par
{\bf 3. $1/q$ expansion, standard approximations, conclusions.}
\par

Let us  expand the  integrand of  the incoherent part,  Eq. (2.5a),  as a
series in  $1/q$.  In order to  enable the integration term  by term, Eq.
(1.1) (and therefore Eq. (2.5))  has to be convoluted with some suitable
smearing  function,  for  instance  the  experimental  resolution.   This
procedure results in a finite asymptotic limit for
$q\to\infty$. In contradistinction the coherent  part
can be shown to vanish exponentially in $q^2$ in that limit
and is totally negligible for finite, large $q$.

One finds for the coefficient functions in the GRS expansion (1.3)

$$\eqalign{F_0(y)&={1\o\sqrt{\pi\beta^2}}\exp (-y^2/\beta^2)\cr
\noalign{\vskip5pt}
(m/q)F_1(y)&=-\l{2\beta\o q}\r \l{y\o {\beta}}\r
\left\(1-{2\o 3}\l{y\o {\beta}}\r^2\right\) F_0(y)\cr
\noalign{\vskip5pt}
((m/q)^2F_2(y)&=-{1\o 6}\l{2\beta\o q}\r^2\left\(1-9\l{y\o {\beta}}\r^2
+8\l{y\o{\beta}}\r^4-{4\o 3}\l{y\o{\beta}}\r^6\right\)F_0(y)\cr
\noalign{\vskip5pt}
(m/q)^3F_3(y)&={1 \o 2}\l{2\beta\o q}\r^3 \l{y\o {\beta}}\r
\left\(1-{47\o 9}\l{y\o{\beta}}\r^2+{74\o{15}}\l{ y\o {\beta}}\r^4
-{4\o 3}\l{y\o {\beta}}\r^6+{8\o 81}\l{y\o{\beta}}\r^8\right\)F_0(y)
}\eqno(3.1)$$
We now compare this result from (2.5)  with
expressions  and approximations which have routinely been used in the
past.  First comes  the GRS  expansion (1.3)  $^3$.  One  checks that
the first  two
coefficients above agree with those  given before $^{8}$.  The expression
for $F_3$ is new and higher order coefficients $F_k$ may be calculated as
well.  The  following comments  are in  order :

\noindent
i) $(m/q)^k\;F_k(y)/F_0(y)$ is a polynomial of order $3k$ in $y/\beta^1$,

\noindent
ii)  the relevant  expansion parameter  is $2\beta/q$,

\noindent
iii) For increasing
$q$ at fixed $y$ the response acquires non-zero contributions from ever
increasing values of the energy $E_n$, with $n\to\infty$ for
$q\to\infty$.
Since for $n\to\infty$ the harmonic oscillator wave function  tends to
a plane wave$^9$, one obtains the standard asymptotic limit $F_0$.

Next we mention the first cumulant approximation $^{10}$ and
an iteration of $F_1$ $^{5,11}$. The
former makes sense only for at least three constituents and we therefore
discuss only the $F_1$ iterant. With $\phi_0(x)=\langle x|0\rangle$, the
ground state wave function
$$\eqalign{\phi(q,y)&=
(2\pi)^{-1}\int^{\infty}_{-\infty} ds e^{iys}\int dx
\phi_0(x-s)\phi_0(x){\rm exp}\left\{(-im/q)\int ^s_0d\sigma
\(V(x-\sigma)-V(x)\)\right\}\cr
&=(2\pi)^{-1}\int^{\infty}_{-\infty} ds e^{iys}{\rm exp}
\(-{1\o 4}{\beta^2s^2}\){\rm exp}\({i\o 6}{\beta^4s^3\o q}\)
{\rm exp}\(-{1\o 4}{\beta^6s^4\o{q^2}}\)\cr
&=(2\pi)^{-1}\int^{\infty}_{-\infty} ds e^{iys}
{\rm exp}\left [-{1\o 4}{\beta}^2s^2\right ] \left \{
1+{1\o 6}i\left [{\beta^4 s^3\o q}\right ] + \Oscr(q^{-2})\right \}
}\eqno(3.2)$$
Note that the coefficient of the term $1/q$ is equal
to the one obtained  from Eq. (2.5a), whereas the other ones are
 quite different.

In conclusion we have applied the Green's function and an
operator technique to obtain the response of
systems interacting through harmonic confining
forces. For those  systems, one can give a closed expression,
Eq. (2.5), which can be tested against approximations$^{5,10,11}$.

The  method can
also applied to other systems  for which the
Green$'$s function in closed form is available, as is the
case for a $\delta$-shell potential
and a particle in an infinitely deep square well$^4$ which has  been
previously investigated, using the summation  method$^1$.

Finally let us note that the form (1.1) shows
the response to be a sum of $\delta$-functions in the energy transfer
$\omega$. Our approach  starts with the everywhere regular Green$'$s
function in the conjugate variable $t$. After performing the
manipulations and integrations as prescribed by (1.2), one obtains
the reduced response $\phi(q,y)$, Eq.(2.5), which is the Fourier
transform of a regular function of $q$ and $t$. On
the one hand all of the above equations are  only valid in the  realm of
the distribution function theory. On the other hand the expansion (1.3)
for the reduced response with the
coefficients (3.1) is a regular function of $q$ and $y$. This apparently
occurs when  $\omega$ is replaced by $y$: the
spacing between the roots of the arguments of the delta functions appearing in
Eq.(1.1)   vanishes for $q\to \infty$.

\par
{\bf Acknowledgements}
\par
One of the authors (ASR) acknowledge useful discussions with
S.A. Gurvitz and M. Kugler.
\noindent

\par

{\bf References}
\bigskip

\refn{$^1$}
S.A. Gurvitz and A.S. Rinat, Phys. Rev. {\bf C47} (1993), to be published

\refn{$^2$}
O.W. Greenberg, Phys. Rev. {\bf C47} (1993) 331.

\refn{$^3$}
H.A. Gersch, L.J. Rodriguez and Phil N. Smith, Phys. Rev. {\bf A5}
(1972) 1547.

\refn{$^4$}
C. Grosche and F. Steiner, SISSA preprints 1/93/FM; 18/93/FM.

\refn{$^5$}
C. Carraro and S.E. Koonin, Phys. Rev. {\bf B41} (1990) 6741;

C. Carraro and S.E. Koonin, Phys. Rev. Lett. {\bf 65} (1990) 2792.

\refn{$^6$}
See for instance L.S. Schulman, Techniques and applications of path
integrations (John Wiley and Sons, NY, 1981).

\refn{$^7$}
See for instance A. Messiah, Quantum mechanics, vol. I, (North
Holland Publishing Company, Amsterdam, 1961).

\refn{$^8$}
A.S. Rinat and R. Rosenfelder, Phys. Lett. {\bf B193} (1987) 411.

\refn{$^{9}$}
I.S. Gradshteyn and I.M. Ryzhik, Table of integrals, series and
products', Academic Press 1980, p. 1034.

\refn{$^{10}$}
H.A. Gersch and  L.J. Rodriguez, Phys. Rev. {\bf A8} (1973) 905.

\refn{$^{11}$}
J. Besprosvany, Phys. Rev. {\bf B43} (1991) 10070.
 \end